\documentclass{aa}
\usepackage{graphicx}
\usepackage{txfonts}
\usepackage{epsfig}
\usepackage{natbib}

\newcommand{\salvo}{}
\newcommand{\refcom}{}
\newcommand{\hd}{HD~81809}   
\newcommand{\acen}{$\alpha$ Cen}   

\newcommand{\lx}{$L_{\rm X}$}  

\newcommand{\xmm}{XMM-\emph{Newton}}

\begin{document} 
\title{The X-ray cycle in the solar-type star  HD\,81809}
\subtitle{\xmm\ observations and implications for the coronal structure}
\author{F.\ Favata\inst{1} 
\and G.\ Micela\inst{2} 
\and S. Orlando\inst{2}
\and J.\,H.\,M.\,M. Schmitt\inst{3} 
\and S.\  Sciortino\inst{2}
\and J.\ Hall\inst{4}
}

\offprints{F. Favata}

\institute{European Space Agency, 8-10 rue Mario Nikis, 75738 Paris cedex 15, France\\
\email{Fabio.Favata@esa.int}  
\and INAF -- Osservatorio Astronomico di Palermo, Piazza del Parlamento 1, I-90134 Palermo, Italy  
  \and  Universit\"at Hamburg, Hamburger Sternwarte, Gojenbergsweg 112, 21029 Hamburg, Germany
  \and Lowell Observatory, 1400 West Mars Hill Road, Flagstaff, Arizona, USA
  }
\date{Received date / Accepted date}

\abstract
{The 11-yr cycle is the best known manifestation of the Sun's activity. While chromospheric cycles have been studied in a number of  solar-like stars, very little is known about how these are reflected in the cyclical behavior of the coronal  X-ray emission in stars other than the Sun.}
{Our long-term \xmm\ program of long-term monitoring of a solar-like star with a well-studied  chromospheric cycle, HD\,81809 aims to study whether an X-ray cycle is present, along with studying its characteristics and its relation to the chromospheric cycle.}
{Regular observations of \hd\ were performed with \xmm, spaced by 6 months from 2001 to 2007. We studied the variations in the resulting coronal luminosity and temperature, and compared them with the chromospheric Ca\,{\sc ii} variations. We also modeled the observations in terms of a mixture of active regions, using a methodology originally developed to study the solar corona. }
{Our observations show a well-defined cycle with an  amplitude exceeding 1 dex and an average luminosity approximately one order of magnitude higher than in the Sun. The behavior of the corona of \hd\ can be modeled well in terms of varying coverage of solar-like active regions, with a larger coverage than for the Sun, showing it to be compatible with a simple extension of the solar case.}
{}


\titlerunning{The X-ray cycle in the solar-type star  HD\,81809}
\authorrunning{F. Favata et~al.}

\maketitle

\section{Introduction}
\label{sec:intro}

Cyclic modulation in the activity level appears to be relatively common among solar-type stars. The best known and studied example is of course the Sun itself, for which the 11-yr cycle was noticed in the form of periodic modulation of the sunspot number already in the mid 19th century. Most activity indicators follow similar cyclical variations (although the study of the Sun's magnetic field shows the true cycle to be 22 yr, with the field polarity reversing each cycle), with the amplitude of the modulation depending strongly on the indicator used. The pioneering and persistent work of O.~Wilson allowed the detection of chromospheric activity cycles on stars other than the Sun, as amply discussed by \cite{bds+95}.

Chromospheric activity cycles appear to be relatively common in intermediate-activity solar type stars, while less active stars are often found in ``Maunder minimum''-like states, i.e.\ showing very little long-term variability (\citealp{bds+95}). The most active stars appear dominated by irregular variability rather than by cyclic behavior. It is, however, not clear if the ``Maunder minimum like'' stars, as well as the Sun itself during the Maunder mininimum, are magnetically less active than the modern solar minimum seems to be, even though they show no HK cycle (\citealp{js2007}). The present-day Sun has a typical chromospheric modulation amplitude, along the 11-yr cycle, of about a factor of 2 (as measured in the $S$ index).

Evidence of cyclical variations in the coronal emission of stars other than the Sun has so far been elusive. In the Sun the X-ray luminosity follows a cycle with the same periodicity as the chromospheric cycle, but with a much larger amplitude, up to a factor of 100 in the Yohkoh 0.73--2.5 keV band (\citealp{act96}). Stars with a high activity level, which have been the typical target of X-ray observations, show modest amounts of long-term variability, with no evidence of a cyclic behavior, as for example reviewed by \cite{ste98b}, and as confirmed from a statistical analysis of the ROSAT All Sky Survey data by \cite{sl2004}. Evidence of long-term X-ray variability with amplitude of a factor of 2.5, and correlated with the chromospheric $S$ index, was reported by \cite{hsb+2003} in 61 Cyg A and B (K5V and K7V) spanning 4.5 years of ROSAT HRI data, suggestive of an activity cycle in X-rays. Some weak statistical evidence of solar-like cycles in lower-activity stars was previously reported by \cite*{hss96} using the RASS observations of the stars with known chromospheric cycles.

To study whether cyclical, high-amplitude variability of the coronal emission is also present on stars other than the Sun, three stellar systems have been monitored since the  beginning of the \xmm\ observatory. With a homogeneous temporal basis of almost 7 years, they represent the best available data set for the study of long-term variability and cycle in stars. The three systems monitored are \hd\ (discussed in the present paper), \acen\ (\citealp{rsf2005}) and 61 Cyg (\citealp{hrs+2006}). Of the three, \hd\ is the one with the longest available observational baseline and the one which shows the clearest evidence to date of a cyclical behavior.

The initial results of the observations of \hd\ were presented by \cite{fmb+2004} (Paper I). High amplitude variability was evident in the data from 2001 to 2003 discussed in Paper I, with some coherence with the known chromospheric cycle of this star. However, no evidence of a coherent minimum was present in the data up to 2003, which made it impossible to distinguish a cyclic behavior from a simpler long-term trend in the star's activity level; the short data set was therefore completely inadequate for any study of the relation between the chromospheric and coronal activity variations. In this paper we present the results from all available observations, spanning 2001 to 2007, from which the cyclical behavior of the star's corona is evident with little if any remaining ambiguity.

\section{Characteristics of \hd}

While considered in the past as a candidate solar twin, \hd\ is in fact a close visual binary with a separation at apoastron of about 0.4 arcsec and a period of about 35 yr (\citealp{pou2000}). The masses of the two components are $M_1 = 1.7 \pm 0.64\, M_\odot$ and $M_2 = 1.0 \pm 0.25\, M_\odot$, with spectral types G2 and G9 and apparent magnitudes $V_1 = 5.8$ and $V_2 = 6.8$ respectively. Both components are slow rotators, with $v\sin i = 3$ km/s (\citealp{sod82}). More details regarding the characteristics of our star can be found in Paper I. \hd\ has more recently been included in the high resolution spectroscopic survey of \cite{tos+2005}, who determined a photospheric Fe abundance of $\rm{[Fe/H]} = -0.34$.

The binary nature of \hd\ does not appear to be a problem for our study, as the very well defined chromospheric cycle reported by \cite{bds+95} shows that the activity-related emission is most likely dominated by one of the two components. Also, the large physical separation of the two stars makes this a non-interacting system as also shown by the low rotational velocity mentioned above.

\begin{figure}
  \begin{center} 
    \leavevmode 
        \epsfig{file=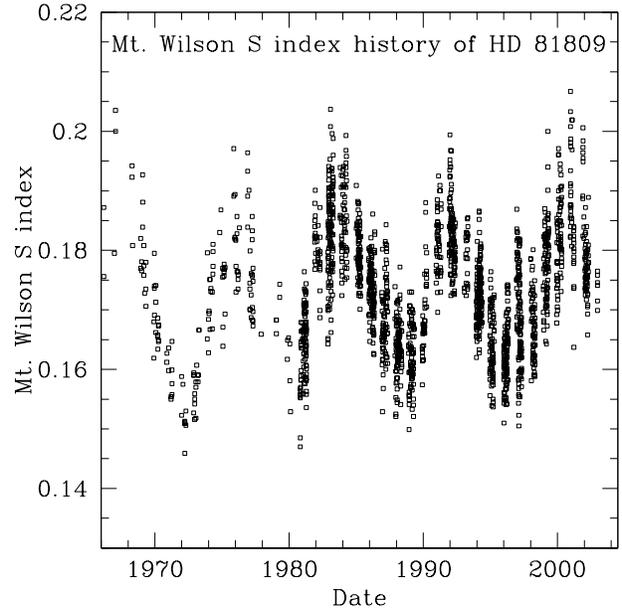, width=8.5 cm}
  \caption{Evolution of the chromospheric activity of \hd\ in the Mt. Wilson $S$ index data,  from 1966 to the end of 2002 (see \citealp{bds+95} for details, data courtesy of S. Baliunas). }
  \label{fig:his}
  \end{center}
\end{figure}

\hd\ is one of the stars in the Mt. Wilson program showing a very clear cyclic behavior with high consistency from one cycle to the other. The chromospheric cycle of \hd\ appears rather similar to the Sun's but with a shorter period of 8.2 yr. The available Mt. Wilson $S$ index observations are plotted in Fig.~\ref{fig:his}. The binary is not resolved, either in the Mt. Wilson data nor in the X-ray observations described below.  

\section{Observations}
\label{sec:obs}

To sample the cyclic behavior of \hd, and considering the chromospheric cycle duration of 8.2 years, we have adopted a strategy based on two observations per year, spaced by 6 months. Given the visibility constraints of the \xmm\ observatory, the observations are executed each year in the Spring and in the Fall (typically in April-May-June and September-October-November). The observations were executed regularly starting in April 2001, with the exception of the Spring 2006 season, in which a technical problem with the scheduling of \xmm\ prevented the observation from being carried out. The length of the observation has been adapted along the program to track the large variation in luminosity shown by the target. Observations up to May 2003 were taken with the medium filter, while the later ones are taken with the thick filter due to a policy change of the observatory.

In addition to the Mt. Wilson data, the Solar-Stellar Spectrograph (SSS) program at Lowell Observatory (Hall \& Lockwood 1995) has accumulated over a decade of Ca II H\&K observations of HD 81809.  The SSS records the HK spectrum from [$\lambda\lambda$] 3860--4010 \AA\ at a resolution of 12\,000, and Hall, Lockwood, \& Skiff (2007) have described the calibration of the spectra to physical flux ($\log R'_{HK}$), and Mt. Wilson $S$.  The raw CCD frames are reduced using an IDL package developed for the SSS data.  Continuing the HD 81809 time series since the termination of the Mt. Wilson observations in 2003, the SSS observations revealed the expected minimum of the well-established 8.2-year cycle in 2004, followed by a steady rise in activity in 2005--2007. In the following, for our comparison between the coronal and chromospheric variability of \hd\ we use this data which has a full temporal overlap with our \xmm\ monitoring program.

The X-ray data presented in this paper have been processed homogeneously using the SAS package. A consistent approach has been adopted in which high background time intervals are first removed from the observation, and then light curves and spectra are extracted. The pn and MOS spectra are simultaneously fit, using {\sc xspec}, with a single temperature optically thin plasma model ({\sc apec}). The metal abundance has been frozen at $Z = 0.3\,Z_\odot$, to reduce the number of free parameters and allow a consistent comparison among the different data sets. The interstellar absorption from the fit is consistent with a zero value.

All observations (with the exception of the June 2002 data set) are well fit with a single-temperature model, with a modest range in best-fit temperature which is however well correlated with the star's X-ray luminosity. To allow a homogeneous analysis of all data sets, a one-temperature fit has been imposed also to the June 2002 data, even though the $\chi^2$ of the fit yields a very low probability. 

\begin{figure}
  \begin{center} 
    \leavevmode 
        \epsfig{file=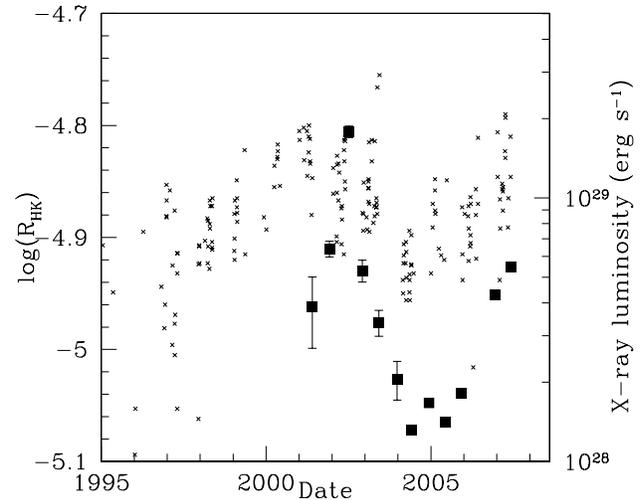, width=8.5 cm}
  \caption{Evolution of the mean (averaged over the duration of each \xmm\ observation, each spanning a few hours) X-ray luminosity in the 0.2--2.5 keV (ROSAT) band (squares, right hand scale) of \hd\ from April 2001 to May 2007 together with the chromospheric Ca \,{\sc ii} activity measured by the SSS at Lowell (crosses, left hand scale). \refcom{For the X-ray data points taken in 2004-2007 the uncertainty on the X-ray luminosity is smaller than the symbol size.}}
  \label{fig:sim}
  \end{center}
\end{figure}

For all observations X-ray luminosities were computed in two bands: a ROSAT-like 0.2--2.5 keV band and for a Yohkoh-like 0.73--3.5 keV band. The resulting spectral parameters for the complete observation timeline are reported in Table~\ref{tab:par}, and a long-term light curve of the X-ray luminosity of \hd\ is plotted (together with the $S$ index) in Fig.~\ref{fig:sim}. None of the observations (with the exception of the June 2002 data point, discussed below) shows visible short-term variability, and each shows a flat light curve.

\begin{table*}[!thbp]
  \begin{center}
    \caption{Best-fit spectral parameters for the 10 \xmm\ observations
      of \hd\ discussed here, spanning 4.5 years.} \leavevmode
    \begin{tabular}{ccccc}
Date & pn rate & \lx\ (0.2--2.5 keV) & \lx\ (0.73--3.5 keV) & $kT$ \\
        & cts s$^{-1}$ & erg s$^{-1}$ & erg s$^{-1}$ & keV \\ \hline
2001-04-25 & $0.1830 \pm 0.055$ & $3.85 \times 10^{28}$ & $1.46 \times 10^{28}$& 0.34 \\
2001-11-01 & $0.3418 \pm 0.024$ & $6.42 \times 10^{28}$ & $2.91 \times 10^{28}$& 0.40 \\
2002-06-06 & $0.7456 \pm 0.037$ & $1.78 \times 10^{29}$ & $8.88 \times 10^{29}$& 0.81 \\
2002-11-02 & $0.2455 \pm 0.024$ & $5.31 \times 10^{28}$ & $2.31 \times 10^{28}$& 0.39 \\
2003-05-03 & $0.1522 \pm 0.017$ & $3.37 \times 10^{28}$ & $1.36 \times 10^{28}$& 0.36 \\
2003-11-22 & $0.0744 \pm 0.012$ & $2.05 \times 10^{28}$ & $7.33 \times 10^{27}$& 0.33 \\
2004-04-30 & $0.0452 \pm 0.002$ & $1.32 \times 10^{28}$ & $4.59 \times 10^{27}$& 0.31 \\
2004-11-09 & $0.0816 \pm 0.003$ & $1.67 \times 10^{28}$ & $6.57 \times 10^{27}$& 0.33 \\
2005-05-03 & $0.0574 \pm 0.002$ & $1.41 \times 10^{28}$ & $4.65 \times 10^{27}$& 0.32 \\
2005-10-30 & $0.0753 \pm 0.002$ & $1.81 \times 10^{28}$& $7.56 \times 10^{27}$& 0.32 \\
2006-11-06 & $0.1993 \pm 0.004$ & $4.28 \times 10^{28}$& $1.71 \times 10^{28}$& 0.36 \\
2007-05-04 & $0.2220 \pm 0.005$ & $5.48 \times 10^{28}$& $2.24 \times 10^{28}$& 0.38\\
    \end{tabular}
    \label{tab:par}
  \end{center}
\end{table*}

The observation of June 2002 is discussed in detail in Paper I. Briefly, although the light curve (which however only spans approximately two hours) shows no evidence of flare-like variability, the spectral characteristics, with a second temperature component at $T_2 = 1.3$ keV and comparable emission measure, suggest a period of enhanced, flare-like activity which drives the X-ray luminosity above the average value for the season. Also, significant short-term variability is present in the June 2002 observation (as e.g.\ evident from Fig.~3 of Paper I), with amplitude of about 30\% in the approximately 2 hours spanned by the observation, indicative of a period of high activity.

As evident from Fig.~\ref{fig:sim}, the first data point (of April 2001) is affected by a much larger uncertainty than the other points. The reason for this (again discussed in detail in Paper I) is a much higher background level across the complete observation, which, coupled with the short duration of the observation, resulted in a high noise level which could not be filtered out.

\section{Discussion}

The available long baseline of X-ray observations of \hd\ for the first time allows us to state that \refcom{\hd\ shows long-term X-ray variability consistent with its regular chromospheric activity cycle}. From a maximum that took place around 2002, the X-ray luminosity decreased to a broad minimum in 2004-2005, and is now rapidly increasing toward a new maximum. The latest observation, taken in May 2007, has an X-ray luminosity very close to the values observed in late 2001, and the maximum amplitude observed,
if one ignores the apparently anomalous June 2002 observation, is about a
factor of 6. If one includes the June 2002 observation, the peak-to-peak
variability amplitude is about a factor of 10 in the ROSAT band and a
factor of about 100 in the Yohkoh-like band.

The coherent behavior of the observed variability clearly shows that one is not in the presence of high-amplitude stochastic variability but rather of a coherent variation in the coronal state of the star, with a time scale of several years, and with an amplitude not dissimilar to the cycle observed in the Sun.

The duration of the chromospheric cycle is well established at 8.2 years. Our X-ray data do not (yet) cover this time span, and therefore we have not yet sampled a complete cycle. The X-ray data points in Fig.~\ref{fig:sim} may give the visual impression of a complete cycle, of a shorter duration, having been covered. However, it is likely that the maximum around 2001 and 2002 has not been properly sampled by our observations: the first point in April 2001 is affected by strong background, and the June 2002 point is likely affected by enhanced, flaring-like activity, resulting in an apparently very sharp maximum which is likely not real. In the Sun the maximum is normally rather broad and
lasts a couple of years, and significant rotational modulation of the X-ray luminosity is present at solar maximum (with an amplitude up to 20\% the total cycle amplitude), so that our single sample around the cycle maximum for \hd\ may be affected by rotation-induced variability. On the other hand the minimum of the cycle is well sampled, and appears broad, with a duration of about 1.5--2 years. A further indication of the fact that the June 2002 point is unlikely to be the true X-ray cycle maximum is the apparent shift between the coronal and chromospheric maximum, as opposed to a good correlation between the chromospheric and coronal minima.

Based on both the observed chromospheric and coronal light curves, the maximum is expected in the next two years. To properly sample the period of maximum coronal activity, and to disentangle the short-term variability which is more likely to be present in periods of enhanced activity, we plan to ask for a more frequent monitoring of the star in
upcoming \xmm\ announcements of opportunity.

\subsection{A solar-like cycle?}

\hd\ is not a strict solar twin, as it is somewhat more massive ($M = 1.7\,M_\odot$) and more evolved, with a radius of $R \simeq 3 R_\odot$. It is also a member of a binary system, with smaller, less massive secondary component. However, the coherent, cyclic variability observed shows that the activity of the system is dominated by only one component in the system, which allows us to ignore the other component for the purpose of the present discussion. We assume in the following that the more massive, larger primary dominates the X-ray activity from \hd, an assumption which will be justified a posteriori as the solar-radius secondary is too small to justify the observed levels of activity with solar-like active regions and filling factors smaller than 1.

\begin{figure}
  \begin{center} 
    \leavevmode 
        \epsfig{file=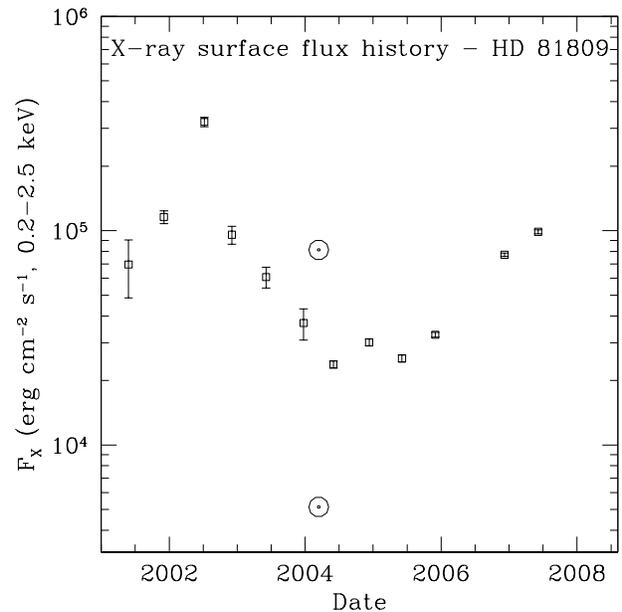, width=8.5cm}
  \caption{Evolution of the X-ray surface flux (in the 0.2--2.5 keV band) of \hd\ from April 2001 to May 2007. The typical X-ray surface flux of the Sun at minimum and maximum of the cycle, in the ROSAT band, is also plotted.}
  \label{fig:lxev}
  \end{center}
\end{figure}

The X-ray luminosity of \hd\ is higher than the Sun's by about 1.5 orders of magnitude. If one however considers the surface flux (assuming that the X-ray emission is dominated by the brighter component in the binary) the overlap with the Sun is large, with maximum flux values (again not considering the June 2002 observation) slightly above the solar ones and minima about mid-way between the solar maximum and minimum values, as shown in Fig.~\ref{fig:lxev}.

To study whether indeed the cycle observed on \hd\ can be explained in terms of a solar-like behavior we have applied the methodology developed in the context of the study of the ``Sun as a star'' \salvo{(\citealt{opr00, porrh00}). As shown by \cite{opr01} from the analysis of a sample of data collected with the Soft X-ray Telescope (SXT; \citealt{tab+91}) on board the Yohkoh satellite,} the evolution of the coronal X-ray luminosity and temperature along the solar cycle can be explained in terms of varying coverage of \salvo{quiet background regions (QR)}, active regions (AR) and cores of active regions (CO), the latter being hotter and brighter than the other components of the \salvo{non-flaring} corona. We have assumed, in our simulation, a minimum radius for the active component (the primary) of \hd\ of $2\,R_\odot$, as this is the more ``challenging'' case (the smaller the star, the higher the filling factors necessary to reproduce the observed level of emission). Therefore the nominal filling factors reported in the following for the various types of coronal structures are to be intended as upper limits to the actual values. In a number of cases we also report the relevant values for the nominal radius of the star $R = 3\,R_\odot$.

The observed temperature and luminosity distribution of the solar corona at minimum can be explained with a surface coverage of AR of about 0.3\%, with no CO present. At solar maximum, the observed emission can be typically explained by a surface
coverage by AR of about 50\% and a further approximately 1\% of CO.
We have explored whether the variability observed for \hd\ along
the cycle can be explained by a mixture of the same ingredients as in the
Sun but in different proportions. To this end, we have considered
the contribution of AR and CO to the luminosity and the spectrum of the
Sun in the X-ray band as derived by \cite{opr04}, studying the long term
evolution of a solar active region observed with Yohkoh/SXT from its
emergence (July 5, 1996) to the decay phase (end of October 1996). Since
the contribution of CO to the X-ray spectrum changes significantly during
the active region evolution\footnote{In particular, \cite{opr04}
have found that the core of active region AR 7978 is characterized
by an average temperature which varies from $\approx 6\times 10^6$ K
during the emergence of the active region (July 5-13, 1996) to $\approx
2\times 10^6$ K at the end of the core evolution circa one month later.},
we have considered the CO during its early evolutionary phase, few days
after the emergence of the active region. 

\begin{figure}
  \begin{center} 
    \leavevmode 
        \epsfig{file=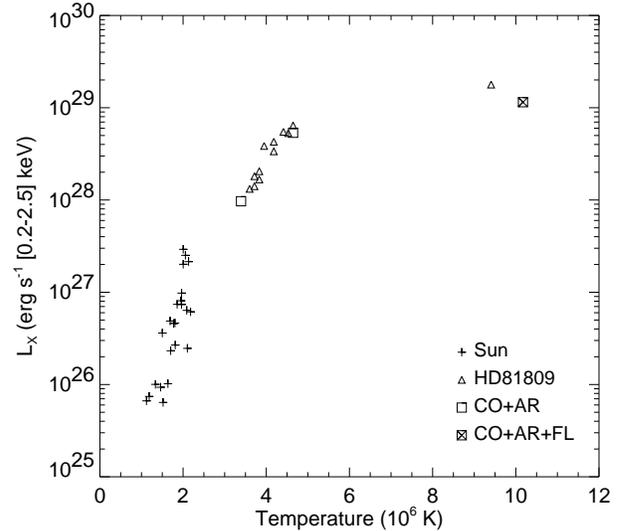, width=8.5cm}
  \caption{The evolution of the coronal X-ray temperature and luminosity along the cycle in both the Sun
           \salvo{(crosses; data adapted from \citealt{opr01}) and \hd\
           (triangles). The figure also shows the synthetic values
           for \hd\ obtained assuming a different coverage of AR and CO
           (squares); the crossed square marks the synthetic values for
           \hd\ considering the contribution of a very intense flare
           (GOES class X9) to the non-flaring corona at the maximum of
           the cycle.}}
  \label{fig:fluxtsim}
  \end{center}
\end{figure}

Then, by simply assuming a permanent coverage of AR with a surface filling factor of about 60\% (27\% for $R = 3\,R_\odot$) and a varying coverage of CO, ranging from about 4\% at the cycle minimum to about 40\% at the cycle maximum (18\% for $R = 3\,R_\odot$), we have \salvo{synthesized the corresponding \xmm\ spectra and} reproduced rather accurately the observed range of variations of \hd\ in terms of both the X-ray temperature and the luminosity. This is shown in Fig.~\ref{fig:fluxtsim}, which shows the observed range of variation in coronal luminosity and temperature for
both the Sun \salvo{(data adapted from \citealt{opr01}) and \hd\ along
their cycle}, together with the synthetic values for \hd\ obtained with
the above approach. \salvo{Table~\ref{tab:solar} reports the resulting
spectral parameters for the synthetic spectra of \hd.}

\salvo{
\begin{table*}[!thbp]
  \begin{center}
    \caption{Best-fit spectral parameters for the synthetic spectra of
      \hd\ derived assuming a different coverage of AR, CO, and flaring
      regions.}
      \leavevmode
    \begin{tabular}{lcccccccccc}
    Phase & \multicolumn{3}{c}{Surface filling factor} &
    $kT_1$ & $kT_2$ & EM$_1$ & EM$_2$ & 
    $\overline{\chi_{\rm r}}^2$ & \lx\ (0.2--2.5 keV) & $kT$ \\
      & AR & CO & FL & [keV] & [keV] & [$10^{50}$ cm$^{-3}$] & [$10^{50}$
     cm$^{-3}$] & & [erg s$^{-1}$] & [keV] \\ \hline
    Max & 60\% & 40\% & - & $0.24\pm 0.03$  & $0.51\pm 0.02$ &
    $2.0\pm 0.4$ & $3.0\pm 0.4$ & 0.65 & $5.35\times 10^{28}$ & 0.40 \\
    Min & 60\% & 4\% & - & $0.16\pm 0.01$  & $0.42\pm 0.03$ &
    $0.51\pm 0.07$ & $0.53\pm 0.08$ & 0.70 & $9.70\times 10^{27}$ & 0.29 \\
    Max + Flare & 60\% & 40\% & 0.0007 & $0.33\pm 0.01$  & $1.26\pm 0.02$ &
    $5.8\pm 0.1$ & $8.1\pm 0.2$ & 0.95 & $1.1\times 10^{29}$ & 0.88 \\
    \hline
    \end{tabular}
    \label{tab:solar}
  \end{center}
\end{table*}
}

\begin{figure}
  \begin{center} 
    \leavevmode 
        \epsfig{file=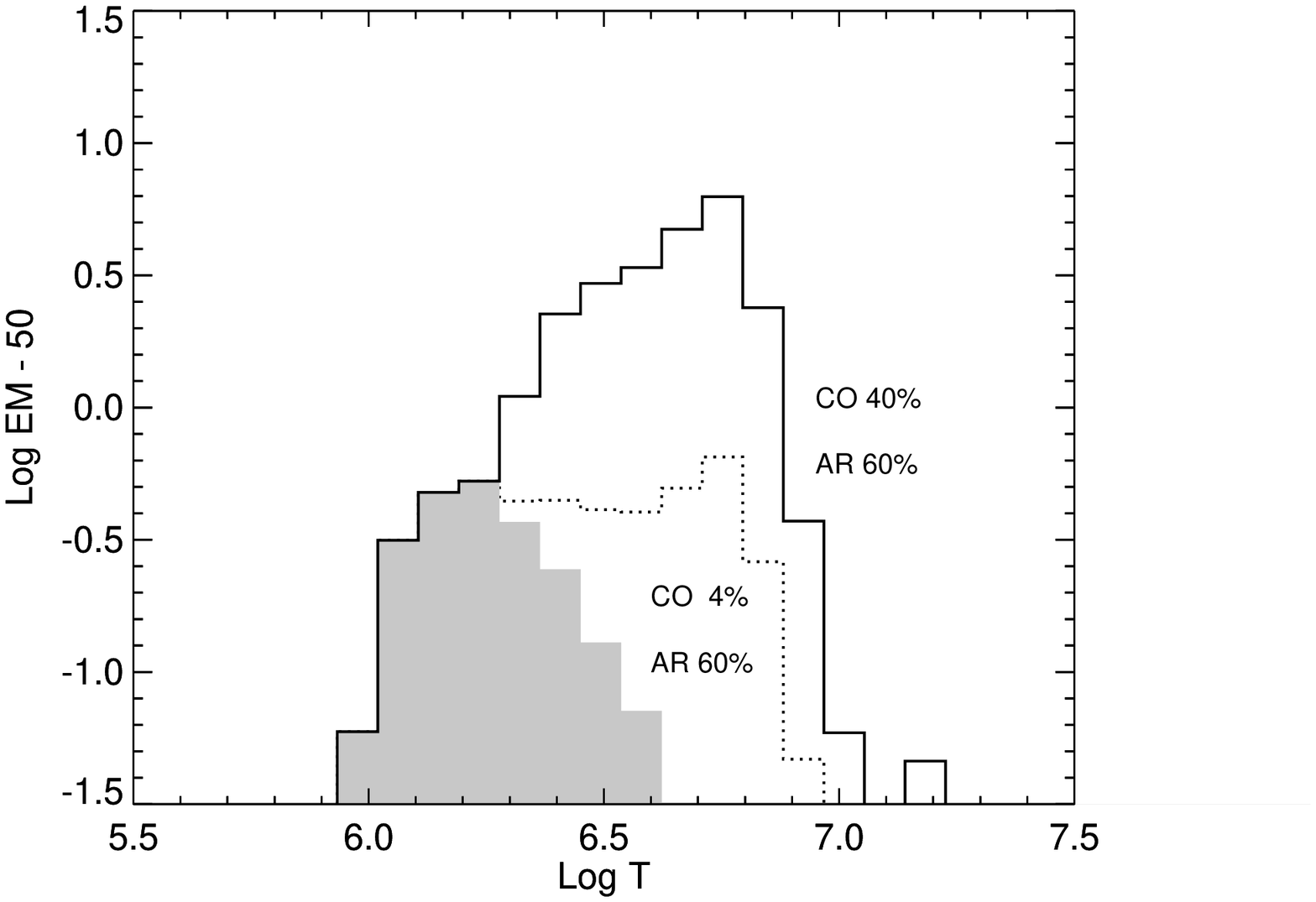, width=8.5cm}
  \caption{The modeled emission measure vs.\ temperature distributions for \hd\ at cycle maximum (solid histogram) and minimum (dotted histogram). The shaded area
 marks the contribution from AR.}
  \label{fig:emt}
  \end{center}
\end{figure}

Fig.~\ref{fig:fluxtsim} shows that while \hd\ is more active, its cyclic behavior lies, in the temperature-luminosity plane, along an extension of the locus occupied by the Sun, and that therefore the phenomenology observed is compatible with being a simple extension of the solar case to a higher activity level. In the case of the Sun, the X-ray variability observed along the cycle is mainly due to the coverage of active regions, whereas the contribution from active region cores is mostly negligible. Our modeling shows that, in the case of \hd\, the variability along the cycle is caused by the active region cores,
which at the maximum of activity for \hd\ cover almost half of the
star. Fig.~\ref{fig:emt} shows the modeled distributions of emission measure vs. temperature, EM$(T)$ at the maximum and minimum of the cycle. Note that the contribution from AR to the EM$(T)$ distribution (and to the synthetic X-ray spectrum) is important close to the minimum of the cycle when it is comparable to the contribution
 from CO. On the other hand, Fig.~\ref{fig:emt} shows that the EM$(T)$
 distribution is largely dominated by CO at the maximum of the cycle,
 whereas the contribution from AR is negligible and the amount of their
 surface filling factor cannot be univocally determined.

Also in the temperature-luminosity diagram the June 2002 point is an
outlier, being well separated from the cloud of values which represent
the variability of \hd\ along the cycle, and therefore likely to be a
separate, flare-like phenomenon. \salvo{This hypothesis is supported
by the fact that adding the contribution of a very intense solar flare (a GOES
class X9 flare observed with Yohkoh/SXT on November 2, 1992) close to
the flare maximum to the synthetic X-ray spectra of \hd\ at the maximum
of its cycle, the synthetic values of X-ray luminosity and average coronal
temperature reproduced are close to those observed (see crossed square
in Fig.~\ref{fig:fluxtsim}).}

The filling factors for the AR and the CO required to explain the observed
variability along the \hd\ cycle are large, but not unrealistic: the AR
coverage is the same as the one observed in the Sun at the cycle maximum,
while the CO coverage at maximum, while much higher than the one observed
in the Sun, is still plausible.

\section{Conclusions}
\label{sec:concl}

We have presented the results from the first 7 years of monitoring of the solar-like star \hd\ (a solar-type star with a well defined cycle in its chromospheric activity) with \xmm, showing strong evidence of cyclic variability with an amplitude of at least 6 in the ROSAT 0.2-2.5 keV band (and even more considering the outlier observed in June 2002). Our long-term \xmm\ program has therefore for the first time resulted in the clear detection of a coronal cycle in a star other than the Sun.

Our observations have however not yet covered a full period of the observed chromospheric cycle. The coverage at the cycle maximum is scarce, and it may be affected by variability of the X-ray emission induced by rotational modulation, so that the single data point does not yet allow us to understand what is the duration of the maximum phase, and to which level is the stochastic variability enhanced with respect to the other phases. The minimum is on the other hand well covered, and it shows a very low level of stochastic variability both in terms of luminosity (a factor $< 2$) and temperature, similar to what is observed in the Sun at the minimum. 

\refcom{At the maximum of activity, our simulations show that the $E\!M(T)$ distribution is dominated by CO (see Fig.~\ref{fig:emt}) which determine the values of average temperature and emission measure reported in Fig.~\ref{fig:fluxtsim}, whereas
the contribution from AR to the total coronal emission is negligible. At the maximum, therefore, only the contribution from CO can be univocally determined, whereas for the AR the quoted value of 60\% is in fact an upper limit to their contribution. When the star is at the minimum of the cycle, the contributions of CO and AR to the $E\!M(T)$ are comparable; changing their relative weight leads to a rapid change in the average temperature. In this case, therefore, the contributions of CO and AR are reasonably constrained: a doubling of the contribution of CO from 4\% to 8\% would lead to an increase in the X-ray luminosity of about 40\%.}

Therefore, even if the absolute activity level of \hd\ at its minimum is much higher than the Sun at minimum, its behavior is very similar. This is somewhat remarkable in that our model shows that a significant fraction of \hd\ at minimum is covered by active regions, unlike the Sun which at minimum is dominated by coronal holes. In the future we plan to study in detail the upcoming maximum, with more frequent observations, to decouple the effect of stochastic variability and of rotational modulation of the X-ray emission.

\hd\ is the subject of a long-term monitoring program performed with the \xmm\ observatory (which also includes $\alpha$ Cen and 61~Cyg). Two more years of observations (at six months cadence) are already planned on the same target, and we plan to re-propose the target in future \xmm\ AOs, to ensure the continuous monitoring. The continuation of the present program will allow us to shed new light on a very basic phenomenon of the Sun, its activity cycles, by studying how similar, somewhat more active stars behave.

\begin{acknowledgements}
  
  GM, SS and SO acknowledge the partial support of the ASI-INAF contracts I/023/05/0, I/015/07/0 and of MUR. JH acknowledges support from the National Science Foundation under grant ATM-0447159. This paper is based on observations obtained with \xmm, an ESA science mission with instruments and contributions directly funded by ESA Member States and the USA (NASA). We would like to thank R.\,A. Donahue as well as the colleagues who have supported the first \xmm\ proposal, namely \salvo{S. Baliunas, A. Collier Cameron, M. G\"udel, F.\,R. Harnden, R. Rosner, R. Stern, K. Strassmeier and F. Walter.}

\end{acknowledgements}

\bibliographystyle{aa}
\bibliography{/Users/Fabio/Documents/Bibliografia/references}

\end{document}